\documentclass[conference]{IEEEtran}
\usepackage{array,cite}
\usepackage{graphicx,epsfig,subfigure}
\usepackage{amsthm}
\usepackage{amsmath}
\usepackage{mdwmath}
\usepackage{array}
\usepackage{subeqnarray}
\usepackage{stfloats}
\usepackage{algorithm,algorithmic}
\usepackage{xcolor}
\usepackage{amssymb}

\newtheorem{theorem}{Theorem}
\newtheorem{lemma}{Lemma}

\begin{document}

\title{Exploiting Delay Correlation for Multi-Antenna-Assisted High Speed Train Communications}

\author{\IEEEauthorblockN{Chunxu~Jiao, Zhaoyang~Zhang$^{\dagger}$, Huazi~Zhang, Liangliang~Zhu}
\IEEEauthorblockA{Department of Information Science and Electronic Engineering, Zhejiang University, Hangzhou 310027, China\\
E-mail: \{jiaocx1990, ning\_ming, hzhang17, zllzju\}@zju.edu.cn}}

\maketitle

\begin{abstract}
In High Speed Train Communications (HSTC), the most challenging issue is coping with the extremely fast fading channel. Compared with its static counterpart, channel estimation on the move consumes excessive energy and spectrum to achieve similar performance. To address this issue, we exploit the \emph{delay correlation} inherent in the linear spatial-temporal structure of multi-antenna array, based on which the rapid fading channel may be approximated by a virtual slow-fading channel. Subsequently, \emph{error probability} and \emph{spectral efficiency} are re-examined for this staticized channel. In particular, we formulate the quantitative tradeoff between the two metrics of interest, by adjusting the \emph{pilot percentage} in each frame. Numerical results verify the good performance of the proposed scheme and elucidate the tradeoff.
\end{abstract}

\section{Introduction}


Recently, there is an emerging research trend towards communication scenarios with high node mobility, such as high speed train and highway communications. To achieve high data-rate in these scenarios, we face multiple challenges, e.g., fast handover, time-varying channel modeling, doubly selective fading modeling, pilot design and channel estimation, etc. However, most of them arise from the ultra-fast channel fading caused by high mobility.


Huge efforts have been made to enhancing wireless communication performance in high mobility scenarios (see \cite{abdi2002space,ist2007deliverable,ma2003optimal,negi1998pilot,barhumi2003optimal,ningsun2014spectral,grossglauser2001mobility,sun2012latency,lu2013novel,zhang2013gossip,sun2011distribution,chen2013efficient,chen2013interference} and the references therein). Among them, \cite{abdi2002space,ist2007deliverable} focus on the mobile channel modeling, based on mathematical analysis and measurement data, respectively. \cite{ma2003optimal,negi1998pilot,barhumi2003optimal,ningsun2014spectral} research into the design of pilot symbols in various mobile systems. Moreover, \cite{grossglauser2001mobility}, being more innovative, takes on node mobility as a type of multiuser diversity, indicating that mobility brings opportunities as well as challenges. However, the intrinsic effect of mobility on communication still needs further study. In mobile communication, will the space-time correlation of the channels be more helpful or more harmful? How can we further mitigate the negative influence caused by the rapid fading channel? How may the MIMO structure affect the communication performance? These problems are of great importance and need for more investigations.

On the one hand, correlation model plays a fundamental element in mobile channel analysis \cite{abdi2002space,gesbert2003theory,molisch2004generic,pedersen2000stochastic}. The space-time modeling framework represents the partial correlation between the channels as well as fast fading and time dispersion. Among the previous works, \cite{abdi2002space} models the Rician fading channel as the combination of diffuse and line-of-sight (LOS) components. Under linear node mobility, there is an inspiring observation that distinct antennas at the same location will share identical CSI. However, the static scattering environment assumption in \cite{abdi2002space} no longer fits in the dynamic scattering considered in our HSTC setting. Therefore, a new spatial-temporal correlation model capturing the dynamic of ambient objects is required for further analysis.

On the other hand, channel estimation using pilot symbols is a fundamental approach for providing robust communication over time-varying channels. However, ultra-high node mobility imposes formidable difficulties in the accurate tracking of channel state information (CSI). A straightforward solution is increasing pilot percentage, which will result in less sub-channels for data transmission and therefore sacrificing the spectral efficiency. In \cite{ma2003optimal} and a recent relevant work \cite{ningsun2014spectral}, pilot is designed to minimize the error probability or maximize the spectral efficiency. Nevertheless, further understanding, especially the quantitative tradeoff between error probability and spectral efficiency, has not been fully revealed in literature.


In this paper, we mainly start with the \emph{delay correlation} phenomenon and an improved channel estimation strategy that takes advantage of the former, and then analyze the tradeoff relations among various metrics of interest (e.g., error probability, spectral efficiency and pilot percentage). Consider the downlink of HSTC that utilizes MIMO techniques and employs pilot-assisted channel estimation. The receive array is usually mounted on the outside of the train and is equipped with multiple antennas. In particular, the receive antennas form a linear array with uniform spacing. Note that the link from base station to the train is an extremely rapid fading channel, which poses a major challenge to high mobility communication. Nevertheless, by exploiting the proposed \emph{delay correlation}, we convert the fast time-varying channels into an approximately static channel. While we focus on the scenario of High Speed Train Communications (HSTC), the results in this paper may apply to general communications with linear node mobility and linear antenna array.

The contributions of this paper are summarized as follows:
\begin{itemize}
    \item The phenomenon of \emph{delay correlation} is discovered. It captures the spatial-temporal channel correlation between antennas moving to the same position at different time.  We formalize this \emph{delay correlation} to facilitate robust mobile communications, especially in HSTCs equipped with linear antenna array.
    \item We propose transmission schemes that exploit delay correlation to convert multiple fast fading channels to a virtual ``static" channel, thus reducing the reliance on excessive pilot insertion. An analytical model is elaborated for this newly established ``staticized" channel.
    \item Based on the ``staticized" channel model, we analyze the tradeoff between error probability and spectral efficiency. This analytical tradeoff provides references for the parameter design in practical systems.
\end{itemize}

The rest of the paper is organized as follows. In Section II, the system model as well as the concept of \emph{delay correlation} is presented. In section III, the impact of pilot percentage on error probability and spectral efficiency is analyzed, after that the tradeoff relation between error probability and spectral efficiency is elucidated. Section IV gives the relevant numerical results. Finally Section V concludes the work.

Throughout the rest of the paper, let $\mathbf{V}\in\mathbb{C}^{M\times N}$ means that the complex matrix $\mathbf{V}$ is consist of $M$ rows and $N$ columns, the capital bold style means it is a matrix and the lowercase bold style means it is a vector. $\textrm{E}\left[\cdot\right]$ is the mathematical expectation operator. $\left ( \cdot  \right )^{*}$, $\left ( \cdot  \right )^{T}$, $\left ( \cdot  \right )^{H}$ stand for complex conjugate, transpose and Hermitian transpose, respectively. $\left\|\cdot\right\|$ refers to the Frobenius norm. $\textrm{trace}\left( \cdot \right)$ is the matrix trace operation.

\section{System Model}

In this section, the concept of \emph{delay correlation} is proposed, based on which the fast fading channel can be approximated by an quasi-static channel. Remarkably, the frames are no longer composed of successive received symbols, but are formed with symbols that benefit from \emph{delay correlation}.

\subsection{Delay Correlation Model}

\emph{Delay correlation} is a unique phenomenon caused by linear mobility and linearly spaced antenna array. To be specific, high speed train has deterministic moving direction and relatively steady speed in a short time. In addition, the receive array, which may utilize massive MIMO techniques, can be positioned to form a line topology in practical design. The aforementioned two assumptions form the foundation for this entire work.

\begin{figure}[!t]\centering
\includegraphics[angle=0,scale=0.35]{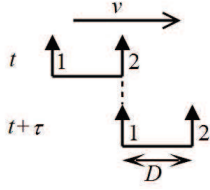}
\caption{A toy example of delay correlation: antenna 1 moves to the same position of antenna 2
after time $\tau$, thus their channels to the source exhibit high correlation.}
\label{Fig_delay_correlation}
\end{figure}

Consider the toy example in Fig. \ref{Fig_delay_correlation}, where two receive antennas move along the line defined by themselves. Then when Antenna 1 moves to the current position of Antenna 2 after time $\tau$, it experiences a similar channel that Antenna 2 experienced $\tau$ time ago\footnote{For simplicity, assume that source S is equipped with single antenna.}. In high speed scenario, $\tau$ can be so short that above mentioned two channels are almost identical. So in this case if Antenna 2 informs Antenna 1 the channel it estimated $\tau$ time ago, then Antenna 1 may save the effort for additional channel estimation. This simple but inspiring phenomenon offers even bigger advantage when the antenna number increases, as we will analyze in detail later.

Regarding correlation modeling, static scattering environment is assumed in previous works, which indicates that the channel will remain unchanged as long as the concerned antennas are immobile. However, if the ambient objects are dynamic, intuitively, there will be a decay factor attached to the cross-correlation coefficient of the diffuse counterpart. In particular, provided with the delay $\tau$ and antenna spacing $D$ for the above example, the correlation coefficient of Antenna 2 at time $t$ and Antenna 1 at time $t+\tau$ is given by the following definition.

\emph{Definition:} The cross-correlation function for mobile frequency non-selective $1\times 2$ Rician fading MIMO channels is
\begin{equation}
\small
\rho ^{\textrm{DIF}} \left ( \tau ,D \right ) = \frac{I_{0}\left( \sqrt{ \kappa^2-\Delta^{2}-j2\kappa \Delta \cos\left( \mu - \gamma \right)} \right)}{\left(K_{R}+1\right)I_{0}\left( \kappa \right)}e^{-c_{0}v\left | \tau \right |},
\label{DIF_Cross-Correlation}
\end{equation}
\begin{equation}
\small
\rho ^{\textrm{LOS}} \left ( \tau ,D \right ) = \frac{K_{R}}{K_{R}+1}e^{j\Delta \cos\left(\gamma\right)},
\label{LOS_Cross-Correlation}
\end{equation}
where $\Delta=2\pi \left(f_{D}\tau - \frac{D}{\lambda}\right)$ presents the location difference of the two antennas, $f_{D}=\frac{v}{\lambda}$ is the maximum Doppler shift and $\lambda$ is the signal wavelength. $\kappa$ controls the width of angle of arrival (AOA) and $\mu\in\left[-\pi, \pi\right)$ accounts for the mean direction of AOA. The mobility of the train is characterized by its velocity $v$ and direction $\gamma$. Finally, $c_{0}$ is a real-valued constant characterizing the inherent spatial property of the scattering objects.

\emph{Remarks:} Some explanations for this concept are in order.

\emph{1)} The cross-correlation of the diffuse part is mainly characterized by three factors: a) location difference; b) AOA difference; c) ambient objects' movement. Apparently, static antennas and static???? scattering environment result in perfect correlation.

\emph{2)} The LOS part only varies in phase of arrival, and the phase shift is obtained easily through geometric analysis.

\emph{3)} This definition is in accordance with the result in \cite{abdi2002space} and may naturally be extended to multiple, even massive, antennas topology.

Based on the above definition, if $\tau$ is set as $\frac{D}{v}$, then $\Delta$ will be $0$, and finally, \emph{delay correlation} is formulated. To be specific, \emph{delay correlation} is characterized by
\begin{equation}
\small
\rho ^{\textrm{DIF}} \left ( \tau ,D \right )|_{\tau=\frac{D}{v}} = \frac{1}{K_{R}+1}e^{-c_{0}D}
\label{Ameliorated_DIF_Cross-Correlation2}
\end{equation}
\begin{equation}
\small
\rho ^{\textrm{LOS}} \left ( \tau ,D \right )|_{\tau=\frac{D}{v}} = \frac{K_{R}}{K_{R}+1}
\label{LOS_Cross-Correlation}
\end{equation}
Note that the direction and velocity of the train are known by both of the base station and the train.

\subsection{Staticized Channel Model}

Inspired by \emph{delay correlation}, the multiple fast-fading channels between the receive antennas and and the base station is converted to a single and slow-fading channel. Therefore, repetitive channel re-estimations becomes dispensable, which will basically resolve the most challenging issue in HSTC.

\begin{figure}[!t]\centering
\includegraphics[angle=0,scale=0.11]{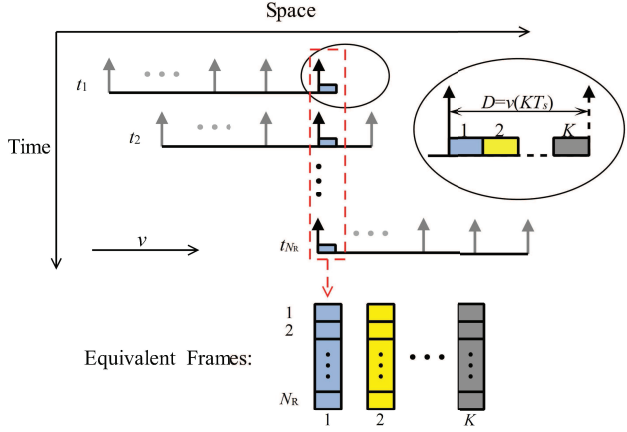}
\caption{Staticized channel model. By exploiting \emph{delay correlation}, fast fading channel can be converted into slow fading channel. Moreover, there will be $K$ equivalent frames received in a staticized block.}
\label{Fig_system_model}
\end{figure}

Fig. \ref{Fig_system_model} shows the staticized channel, which is a virtualized static channel between the source antenna and different receive antenna at different time. In particular, assume that the receiver is equipped with $N_{\textrm{R}}$ antennas. Since base station is informed of receive array spacing $D$ and velocity of the train $v$, it is reasonable to assume that base station being able to adjust the symbol time $T_{s}$ so that CSI will not change in $T_{s}$ and there exists an integer $K$ which satisfies $D=v\left ( KT_{s} \right )$. By utilizing antenna selection, one antenna of the array is activated each time, marked as the dark ones in Fig. \ref{Fig_system_model}, to process the transmit signal. As the CSI variation of the selected antennas is negligible, channel re-estimations can be reduced in $N_{\textrm{R}}$ time slots.

Assume Rician fading channels between $\textrm{S}$ and the selected antennas $h_{i}\in \mathbb{C},i=1,\dots,N_{\textrm{R}}$. Then the signals observed at the train is given by
\begin{equation}
\small
y_{i}=\sqrt{E_{0}}h_{i}x_{i}+n_{i},~i=1,\dots,N_{\textrm{R}}
\label{Received-Signal}
\end{equation}
where $y_{i}\in \mathbb{C}$ is the received signal, $n_{i}\in \mathbb{C}$ is the zero-mean complex Gaussian noise with covariance coefficient $\sigma ^{2}_{n}$, $E_{0}$ is the average transmission energy of a symbol, scalar $x_{i}$ is the transmitted signal. To estimate the channel, several $x_{i}$ are selected to act as pilot symbols.


$\mathbf{h}=\left[ h_{1}~h_{2}~\dots~h_{N_{\textrm{R}}}\right]^{T}$ is an equivalent slow fading channel. Utilizing the aforementioned \emph{delay correlation} coefficients (\ref{Ameliorated_DIF_Cross-Correlation2}) and (\ref{LOS_Cross-Correlation}), the cross-correlation between $h_{p}$ and $h_{q}$, $1\leq p<q\leq N_{\textrm{R}}$, is given by
\begin{equation}
\small
\rho ^{\textrm{DIF}}_{p,q} \left ( \tau ,\left ( q-p \right )D \right )|_{\tau=\frac{\left ( q-p \right )D}{v}}= \frac{1}{K_{R}+1}e^{-c_{0}\left ( q-p \right )D},
\label{DIF_Delay-Correlation}
\end{equation}
\begin{equation}
\small
\rho ^{\textrm{LOS}}_{p,q} \left ( \tau ,\left ( q-p \right )D \right )|_{\tau=\frac{\left ( q-p \right )D}{v}}= \frac{K_{R}}{K_{R}+1},
\label{LOS_Delay-Correlation}
\end{equation}

$\mathbf{x}=\left[ x_{1}~x_{2}~\dots~x_{N_{\textrm{R}}}\right]^{T}$ is a vertical equivalent frame which involves $N_{\textrm{R}}$ symbols. It is noteworthy that the downlink transmission is actually divided into many successive staticized blocks, each lasts for a period of $N_{\textrm{R}}\times K T_{s}$, so there will be $K$ frames received in a block as depicted in Fig.\ref{Fig_system_model}.

With $\mathbf{h}$ and $\mathbf{x}$ defined above, we formalize the system model as
\begin{equation}
\small
\mathbf{y}=\sqrt{E_{0}}\mathbf{X}\mathbf{h}+\mathbf{n},
\label{Equivalent-Received-Signal}
\end{equation}
where $\mathbf{y}=\left[ y_{1}~y_{2}~\dots~y_{N_{\textrm{R}}}\right]^{T}\in \mathbb{C}^{N_{\textrm{R}}\times 1}$ is the received vertical vector, $\mathbf{n}=\left[ n_{1}~n_{2}~\dots~n_{N_{\textrm{R}}}\right]^{T}\in \mathbb{C}^{N_{\textrm{R}}\times 1}$ is the vertical noise vector with covariance matrix $\mathbf{R}_{n}=\sigma ^{2}_{n}\mathbf{I}_{N_{\textrm{R}}}$, $\mathbf{X}$ is a size-$N_{\textrm{R}}$ diagonal matrix with $\mathbf{x}$ on its diagonal.

\section{Tradeoff between Error Probability and Spectral Efficiency}

Based on the staticized channel model derived in Section II, there are $K$ equivalent vertical frames in a staticized
block, each of the frames goes through a highly correlated channel environment. To guarantee the communication performance, several symbols serve as pilots, inducing spectral efficiency loss, though. Pilots should be sufficient, but not excessive, so it is important to determine the necessary pilot percentage when there is a performance requirement. In this sense, a tradeoff between error probability and spectral efficiency is formulated.

\subsection{Channel Estimation}

The \emph{Two-Step MMSE} scheme proposed in \cite{wu2012optimum} is employed for channel estimation. It decomposes the channel estimation into two steps, i.e., channel estimation at pilot locations and channel interpolation. With similar definitions, $N_{\textrm{R}}$ time slots are divided into $N_{s}$ data symbols and $N_{p}\le N_{s}$ pilot symbols. $N_{\textrm{R}}$, $N_{s}$ and $N_{p}$ can be chosen such that $L=\frac{N_{s}}{N_{p}}$ is an integer. The pilot symbols are equally spaced such that two adjacent pilot symbols are interpolated by $L$ data symbols. In this sense, the pilot percentage is $\delta=\frac{1}{L+1}$. Assume that the stationary LOS part of Rician fading channel $h^{\textrm{LOS}}$ is perfectly estimated, which implies that channel estimation will only apply to the diffuse component $h^{\textrm{DIF}}$. Hence, using the model in (\ref{Received-Signal}), the mean square error (MSE) of the $i-$th channel coefficient is
\begin{equation}
\small
\sigma_{i}^{2}=\textrm{E}\left[ \left| \hat{h}_{i}-h_{i} \right|^{2} \right]=\textrm{E}\left[ \left| \hat{h}_{i}^{\textrm{DIF}} -h_{i}^{\textrm{DIF}} \right|^{2} \right],~i=1,\dots,N_{\textrm{R}},
\label{MSE}
\end{equation}
where $\hat{h}_{i}$ is an estimation of $h_{i}$.

Firstly, the receiver obtains the channel estimations at pilot locations. Assume that $\mathbf{P}$, $\mathbf{h}_{p}$ and $\mathbf{y}_{p}$ are the pilot symbols and their corresponding fading channels, received symbols extracted from $\mathbf{X}$, $\mathbf{h}$ and $\mathbf{y}$, respectively. Then the received pilot symbols are expressed as
\begin{equation}
\small
\mathbf{y}_{p}=\sqrt{E_{0}}\mathbf{P}\mathbf{h}_{p}+\mathbf{n}_{p},
\label{Received-Pilots}
\end{equation}
The receive array derives the channel estimations by minimizing the average MSE, $\sigma_{\textrm{avg}}^{2}=\frac{1}{N_{p}}\textrm{E}[ \| \hat{\mathbf{h}}_{p}^{\textrm{DIF}}-\mathbf{h}_{p}^{\textrm{DIF}} \|^{2} ]$. Its solution is
\begin{equation}
\small
\hat{\mathbf{h}}_{p}^{\textrm{DIF}}=\mathbf{W}_{p}^{H} \left( \mathbf{y}_{p} - \sqrt{E_{0}}\mathbf{P}\mathbf{h}_{p}^{\textrm{LOS}}  \right),
\label{Pilot-Location-Estimation}
\end{equation}
where $\hat{\mathbf{h}}_{p}^{\textrm{DIF}}\in \mathbb{C}^{N_{p}\times 1}$ is the estimation of $\mathbf{h}_{p}^{\textrm{DIF}}$, and $\mathbf{W}_{p}=\sqrt{E_{0}} ( E_{0}\mathbf{P}\mathbf{R}_{hh}^{\textrm{DIF}}\mathbf{P}^{H}+ \sigma ^{2}_{n}\mathbf{I}_{N_{p}})^{-1}\mathbf{P}\mathbf{R}_{hh}^{\textrm{DIF}}$ is the channel MMSE estimation matrix, auto-correlation matrix of the diffuse channel $\mathbf{R}_{hh}^{\textrm{DIF}}=\textrm{E}\left[ \mathbf{h}_{p}^{\textrm{DIF}}\left(\mathbf{h}_{p}^{\textrm{DIF}}\right)^{H} \right]\in \mathbb{R}^{N_{p}\times N_{p}}$ is a Toeplitz matrix whose $\left( m,n \right)$-th element being
\begin{equation}
\small
\rho_{m,n}=\rho_{m,n}^{\textrm{DIF}}= \frac{1}{K_{R}+1}e^{-c_{0}\left | m-n \right |\frac{D}{\delta}},
\label{Rhh-Auto-Correlation}
\end{equation}
where $\delta$ is the pilot percentage.

In the second step, the channel estimation of arbitrary $h_{i}$ is obtained through interpolating $\hat{\mathbf{h}}_{p}^{\textrm{DIF}}$ and attaching the LOS component
\begin{equation}
\small
\hat{h}_{i}=\mathbf{w}_{d}^{H}\hat{\mathbf{h}}_{p}^{\textrm{DIF}}+h_{i}^{\textrm{LOS}},
\label{Interpolation}
\end{equation}
where the real-valued coefficient vector $\mathbf{w}_{d}\in \mathbb{R}^{N_{p}\times 1}$ is designed according to the MSE minimization criterion (\ref{MSE}).

It is proved in \cite{wu2012optimum} that this scheme is equivalent to the optimum linear MMSE estimator as follows
\begin{equation}
\small
\begin{split}
\hat{h}_{i}=&\sqrt{E_{0}} \mathbf{r}_{i}^{\textrm{DIF}}\mathbf{P}^{H} ( E_{0}\mathbf{P}\mathbf{R}_{hh}^{\textrm{DIF}}\mathbf{P}^{H}+ \sigma ^{2}_{n}\mathbf{I}_{N_{p}})^{-1} \times \\
&\left(\mathbf{y}_{p} - \sqrt{E_{0}}\mathbf{P}\mathbf{h}_{p}^{\textrm{LOS}} \right)+h_{i}^{\textrm{LOS}},~i=1,\dots,N_{\textrm{R}},
\label{MMSE-Estimation}
\end{split}
\end{equation}
where $\mathbf{r}_{i}^{\textrm{DIF}}=\textrm{E}\left[ h_{i}^{\textrm{DIF}} \left( \mathbf{h}_{p}^{\textrm{DIF}} \right)^{H} \right]$, which indicates that $\mathbf{w}_{d}^{H}=\mathbf{r}_{i}^{\textrm{DIF}}\left(\mathbf{R}_{hh}^{\textrm{DIF}}\right)^{-1}$.

\subsection{Error Probability vs. Pilot Percentage}

The accuracy of channel estimation is mainly determined by pilot percentage. Meanwhile, imperfect channel estimation results in high error probability\cite{gifford2005diversity}. Hence, there exists a mapping between error probability $P_{e}$ and pilot percentage $\delta$.

Above all, the MSE of the \emph{Two-Step MMSE} estimation scheme is derived. As stated in \cite{ningsun2014spectral}, define the error correlation matrix at pilot locations as
\begin{equation}
\small
\mathbf{R}_{ee}=\textrm{E}\left[ \mathbf{e}_{p}\mathbf{e}_{p}^{H} \right],
\label{Error-Correlation}
\end{equation}
where $\mathbf{e}_p=\hat{\mathbf{h}}_p-\mathbf{h}_p=\hat{\mathbf{h}}_{p}^{\textrm{DIF}}-\mathbf{h}_{p}^{\textrm{DIF}}$. For simplicity, assume that the pilot symbols $\mathbf{P}=\mathbf{I}_{N_{p}}$. Substituting this assumption and (\ref{Received-Pilots})(\ref{Pilot-Location-Estimation}) into (\ref{Error-Correlation}), it can be calculated that
\begin{equation}
\small
\mathbf{R}_{ee}=\mathbf{R}_{hh}^{\textrm{DIF}}-\mathbf{R}_{hh}^{\textrm{DIF}}\left( \mathbf{R}_{hh}^{\textrm{DIF}}+\frac{1}{\gamma}\mathbf{I}_{N_{p}} \right)^{-1}\mathbf{R}_{hh}^{\textrm{DIF}},
\label{Pilot-Error-Correlation}
\end{equation}
where $\gamma=\frac{E_{0}}{\sigma_{n}^{2}}$ is the signal-to-noise ratio (SNR). Then the average MSE at pilot locations is
\begin{equation}
\small
\sigma_{p,N_{p}}^{2}=\frac{1}{N_{p}}\textrm{trace}\left( \mathbf{R}_{ee} \right),
\label{MSE-Pilot-Locations}
\end{equation}
Through asymptotic analysis, i.e., $N_{p}\rightarrow\infty$, $N_{s}\rightarrow\infty$ and $N_{\textrm{R}}\rightarrow\infty$, with a finite pilot percentage $\delta$, the asymptotic MSE at pilot locations can be expressed as a function of $\delta$ with the following theorem. Note that asymptotic analysis is reasonable for massive MIMO where a large amount of receiver antennas exist.

\begin{theorem}
Let $N_{p}\rightarrow\infty$ and $N_{s}\rightarrow\infty$ while keeping a finite pilot percentage $\delta$, the asymptotic MSE $\sigma_{p}^{2}=\lim_{N_{p}\rightarrow\infty}\sigma_{p,N_{p}}$ at pilot locations is
\begin{equation}
\small
\sigma_{p}^{2}=\sqrt{\frac{1}{\gamma^{2}+2\gamma\frac{1+\alpha^2}{1-\alpha^2}\left(K_{R}+1\right)+\left(K_{R}+1\right)^{2}}},
\label{Theorem1}
\end{equation}
where $\alpha=e^{-c_{0}\frac{D}{\delta}}$.
\end{theorem}

\begin{IEEEproof}[Proof]
The proof is in Appendix A.
\end{IEEEproof}

\emph{Remarks:} We can see that the estimation MSE will decrease when SNR and Rician factor rise. And the increase in pilot percentage leads to higher correlation between channels at adjacent pilot locations. Moreover, stronger intraclass correlation facilitates more accurate estimations. Therefore, increase in pilot percentage will concurrently result in smaller MSE. These observations will be further verified through the simulation results.

Next, the MSE of arbitrary $h_{i}^{\text{DIF}}$ needs to be calculated. Decompose the remaining $N_{s}$ data symbols into $L$ groups, each with $N_{p}$ symbols.

Define the $u$-th symbol group as $\mathbf{X}_{u}\in\mathbb{C}^{N_{p}\times N_{p}}$, which contains the data symbols with indices $\left\{ i'_{k} \right\}=\left( k-1 \right)\left(L+1\right)+u$, $k=1,\dots,N_{p}$. In addition, define its corresponding channels and received symbols as $\mathbf{h}_{d,u}$ and $\mathbf{y}_{d,u}$, respectively. In this case, the system model becomes
\begin{equation}
\small
\mathbf{y}_{d,u}=\sqrt{E_{0}}\mathbf{X}_{u}\mathbf{h}_{d,u}+\mathbf{n}_{u},
\label{Data-System-Model}
\end{equation}
From (\ref{Interpolation}), the MMSE estimation of $\mathbf{h}_{d,u}$ is
\begin{equation}
\small
\hat{\mathbf{h}}_{d,u}=\mathbf{W}_{d,u}^{H}\hat{\mathbf{h}}_{p}^{\textrm{DIF}}+\mathbf{h}_{d,u}^{\textrm{LOS}},
\label{Data-MMSE-Estimation}
\end{equation}
where $\mathbf{W}_{d,u}^{H}=\mathbf{R}_{dh,u}^{\textrm{DIF}}\left( \mathbf{R}_{hh}^{\textrm{DIF}} \right)^{-1}$ contains the interpolating coefficients, and $\mathbf{R}_{dh,u}^{\textrm{DIF}}=\textrm{E}\left[ \mathbf{h}_{d,u}^{\textrm{DIF}}\left(\mathbf{h}_{p}^{\textrm{DIF}} \right)^{H} \right]\in\mathbb{R}^{N{p}\times N_{p}}$ is also a Toeplitz matrix whose $(m,n)$-th element is
\begin{equation}
\small
\rho_{m,n}=\rho_{m,n}^{\textrm{DIF}}= \frac{1}{K_{R}+1}e^{-c_{0}\left | m-n +u\delta \right |\frac{D}{\delta}},
\label{Rdh-Auto-Correlation}
\end{equation}
Then the corresponding error correlation matrix of channel estimations at these symbol locations, defined as $\mathbf{\Psi}_{ee,u}=\textrm{E}[ \mathbf{e}_{d,u}\mathbf{e}_{d,u}^{H} ]$ with $\mathbf{e}_{d,u}=\hat{\mathbf{h}}_{d,u}-\mathbf{h}_{d,u}=\hat{\mathbf{h}}_{d,u}^{\textrm{DIF}}-\mathbf{h}_{d,u}^{\textrm{DIF}}$, can be calculated as
\begin{equation}
\small
\mathbf{\Psi}_{ee,u}=\mathbf{R}_{dd,u}^{\textrm{DIF}}-\mathbf{R}_{dh,u}^{\textrm{DIF}}\left( \mathbf{R}_{hh}^{\textrm{DIF}} +\frac{1}{\gamma}\mathbf{I}_{N_{p}}\right)^{-1}\left(\mathbf{R}_{dh,u}^{\textrm{DIF}}\right)^{H},
\label{Data-Error-Correlation}
\end{equation}
where $\mathbf{R}_{dd,u}^{\textrm{DIF}}=\Big[ \mathbf{h}_{d,u}^{\textrm{DIF}} \left(\mathbf{h}_{d,u}^{\textrm{DIF}}\right)^{H} \Big]=\mathbf{R}_{hh}^{\textrm{DIF}}$. Similarly, the average MSE for the channel estimations at $u$-th group data symbol locations is
\begin{equation}
\small
\sigma_{d,u,N_{p}}^{2}=\frac{1}{N_{p}}\textrm{trace}\left( \mathbf{\Psi}_{ee,u} \right),
\label{MSE-Data-Locations}
\end{equation}
Through asymptotic analysis, the MSE at data symbol locations is given by the following theorem.
\begin{theorem}
Let $N_{p}\rightarrow\infty$ and $N_{s}\rightarrow\infty$ while keeping a finite pilot percentage $\delta$, the asymptotic MSE $\sigma_{d,u}^{2}=\lim_{N_{p}\rightarrow\infty}\sigma_{d,u,N_{p}}$ at $u$-th group data symbol locations is
\begin{equation}
\small
\sigma_{d,u}^{2}=\sigma_{p}^{2}+\frac{1}{2\pi}\int_{-\pi}^{\pi}
\left [  \frac{  \Lambda\left ( \Omega  \right )^{2} - \left|\Lambda_{dh,u}\left ( \Omega  \right )\right|^{2}  }{\Lambda\left ( \Omega  \right )  + \frac{1}{\gamma}} \right ]d\Omega,
\label{Theorem2}
\end{equation}
where $\Lambda\left ( \Omega  \right )$ and $\Lambda_{dh,u}\left ( \Omega  \right )$ are as follows
\begin{equation}
\small
\Lambda\left ( \Omega  \right )=\frac{1}{K_{R}+1}\left[ \frac{1-\alpha^{2}}{ 1- 2\alpha\textrm{cos}(\Omega)+\alpha^{2} }\right],
\label{Theorem2-1}
\end{equation}
\begin{equation}
\small
\Lambda_{dh,u}\left ( \Omega  \right )=\frac{1}{K_{R}+1}\left[ \frac{\alpha\left( \beta^{-1}-\beta \right) e^{j\Omega} + \beta-\alpha^{2}\beta^{-1}}{ 1- 2\alpha\textrm{cos}(\Omega)+\alpha^{2} }\right],
\label{Theorem2-2}
\end{equation}
$\alpha=e^{-c_{0}\frac{D}{\delta}}$ and $\beta=e^{-c_{0}uD}$.
\end{theorem}

\begin{IEEEproof}[Proof]
The proof is in Appendix B.
\end{IEEEproof}


\emph{Remarks:} In low SNR regime, the estimation error will be mainly dominated by noise, while $\beta$, the correlation coefficient between the channels of the $u$-th symbol group and the prior pilot symbols, has negligible effect on the MSE. In this case, the second item of the expression would be very close to $0$, indicating that the MSE at data locations is quite similar to that at pilot locations. However, in high SNR regime, channels at pilot locations are perfectly estimated, so noise will no longer affect the estimation error while \emph{delay correlation} becomes more predominating. In this sense, the first item approaches $0$ and the second item becomes almost a positive constant, along which the diversity order would be zero.

Having obtained the analytical MSEs, it is possible to analyze the error probability of our system model. For BPSK, error probability $P_{e}$ is equal to BER. The maximum likelihood decision rule is
\begin{equation}
\small
\hat{x}_{i}=\mathop{\arg\min}_{x\in\left\{-1,1 \right\}} \left\{ \left| \frac{1}{\sqrt{E_{0}}}h_{i}^{*}y_{i}-x \right| \right\},
\label{Decision}
\end{equation}
Thus the BER performance is given in\cite{wu2008optimal}
\begin{equation}
\small
\begin{split}
P_{e,u}=&\frac{1}{\pi}e^{-\frac{K_{R}}{\rho^{2}}}\int_{0}^{\frac{\pi}{2}}\left[ 1+\frac{\tilde{\gamma }}{K_{R}+1}\frac{1}{\textrm{sin}^{2}\left ( \phi \right )} \right]^{-1} \times\\
&\textrm{exp}\left \{ \frac{K_{R}}{\rho^{2}} \left[ 1+\frac{\tilde{\gamma }}{K_{R}+1}\frac{1}{\textrm{sin}^{2}\left ( \phi \right )} \right]^{-1}\right \}
d\phi,
\label{BER}
\end{split}
\end{equation}
where $\rho=\sqrt{\frac{1}{1+(K_{R}+1)\sigma_{d,u}^{2}}}$ is the covariance coefficient between $\mathbf{h}_{d,u}$ and $\hat{\mathbf{h}}_{d,u}$, $\tilde{\gamma }=\frac{\left( K_{R}+1 \right)\rho^{2}}{\gamma \left( 1-\rho^{2} \right)+K_{R}+1}\gamma=\frac{1}{\sigma_{d,u}^{2}+\frac{1}{\gamma}}$ is the equivalent SNR.

As $\sigma_{d,u}^{2}$ is related to the index $u$, the average BER is simply derived by calculating the mean of $P_{e,u}$ over $u$
\begin{equation}
\small
P_{e}= \textrm{E}_{u}\left[P_{e,u}\right]=\frac{1}{L}\sum_{u=1}^{L}P_{e,u}.
\label{Avg-BER}
\end{equation}

\subsection{Spectral Efficiency vs. Pilot Percentage}

Intuitively, more pilot symbols result in better channel estimations, which would help improve the spectral efficiency. On the other hand, excessive pilot symbols would lead to unnecessary spectrum overhead. Hence, there exists an optimal pilot percentage which would maximize the spectral efficiency. \cite{ningsun2014spectral} gives the maximum spectral efficiency analysis with imperfect channel information in Rayleigh fading channels. In a similar fashion, the relation between spectral efficiency $\eta$ and pilot percentage $\delta$ in Rician fading models will be obtained.

In this paper, the effective spectral efficiency is defined as
\begin{equation}
\small
\eta=\frac{N_{s}}{N_{\textrm{R}}}\textrm{E}_{\hat{h}}\left[ C\left( \hat{h},\gamma \right) \right]=\left( 1-\delta \right)\textrm{E}_{\hat{h}}\left[ C\left( \hat{h},\gamma \right) \right],
\label{Effecive-Spectrum-Efficency}
\end{equation}
where $C( \hat{h},\gamma )$ is the channel capacity with imperfect channel estimations $\hat{h}$ and SNR $\gamma$.

\begin{lemma}
For a M-PSK modulation system operating in a Rician fading channel with pilot-assisted MMSE channel estimation with regard to the $u$-th symbol group, the channel capacity conditioned on imperfect channel estimation $\hat{h}$ and SNR $\gamma$ is upper bounded by
\begin{equation}
\small
C( \hat{h},\gamma )_{u,up}=\log\left( 1+| \hat{h} |^2\tilde{\gamma } \right),
\label{Capacity-Imperfect-Estamition}
\end{equation}
where $\tilde{\gamma }=\frac{1}{\sigma_{d,u}^{2}+\frac{1}{\gamma}}$ is the equivalent SNR.
\end{lemma}

\begin{IEEEproof}
The proof is in Appendix C.
\end{IEEEproof}

Based on the above lemma, take the average of $C( \hat{h},\gamma )_{u,up}$ over $\hat{h}$ and $u$, and then substitute the expectation into (\ref{Effecive-Spectrum-Efficency}), the expression of effective spectral efficiency is derived.

\begin{theorem}
For the system operating in a Rician fading channel with pilot-assisted MMSE channel estimation with regard to the $u$-th symbol group, the effective spectral efficiency is given by
\begin{equation}
\small
\eta_{u,up}\approx\left( 1-\delta \right)\int_{0}^{+\infty }C\left ( x \right )p\left ( x \right )dx,
\label{Theorem3}
\end{equation}
where $C(x)$ and $p(x)$ is expressed as
\begin{equation}
\small
C\left ( x \right )=\log\left( 1+\frac{x}{2\left(K_{R}+1\right)}\tilde{\gamma }\right),
\label{Theorem3-1}
\end{equation}
\begin{equation}
\small
p\left ( x \right )=\frac{1}{2}e^{-\frac{x+2K_{R}}{2}}I_{0}\left ( \sqrt{2K_{R}x} \right ),
\label{Theorem3-2}
\end{equation}
and $\tilde{\gamma }=\frac{1}{\sigma_{d,u}^{2}+\frac{1}{\gamma}}$ is the equivalent SNR.
\end{theorem}

\begin{IEEEproof}
The proof is in Appendix D.
\end{IEEEproof}

\emph{Remarks:} Apparently, the spectrum is used more efficiently in high SNR regime. Furthermore, as the pilot percentage increases, data percentage will keep decreasing while the corresponding capacity of the data channel maintains rising. Note that the effective spectral efficiency is the product of data percentage and the average data channel capacity, therefore, there exists an optimal pilot percentage that maximizes the spectral efficiency.

Similarly, the spectral efficiency should be averaged on $u$, thus getting the final result
\begin{equation}
\small
\eta_{up}=\frac{1}{L}\sum_{u=1}^{L}\eta_{u,up}.
\label{Avg-SE}
\end{equation}

\subsection{Error Probability vs. Spectral Efficiency}

After SNR $\gamma$ and pilot percentage $\delta$ are specified, the corresponding MSE at data symbol locations can be calculated through (\ref{Theorem2}). Furthermore, the average error probability $P_{e}$ and spectral efficiency $\eta_{up}$ can be derived through (\ref{Avg-BER}) and (\ref{Avg-SE}), respectively. Evidently, it is unrealistic to simultaneously achieve lower $P_{e}$ and higher $\eta_{up}$, thus implying a tradeoff relation between them.

According to (\ref{Theorem3}), it is possible to obtain the maximum spectral efficiency through selecting a specific $\delta_{\textrm{SE-opt}}$, yet the corresponding BER performance would not necessarily be satisfactory. However, it is worth nothing that $\delta_{\textrm{SE-opt}}$ has practical implication in our tradeoff analysis, i.e., it would serve as the lower bound of feasible pilot percentages. When $\delta>\delta_{\textrm{SE-opt}}$, both spectral efficiency and error probability would decrease monotonically. Hence, it is crucial to choose a pilot percentage between $\delta_{\textrm{SE-opt}}$ and $0.5$ to balance between error probability and spectral efficiency.

Having understood the relation among MSE, error probability and spectral efficiency, it makes sense to adjust $\delta$ between $\delta_{\textrm{SE-opt}}$ and $0.5$ while calculating the corresponding $P_{e}$ and $\eta_{up}$, thus obtaining the optimal tradeoff between error probability and spectral efficiency.

\section{Numerical Results}
Numerical results are provided in this section. Firstly, the impacts of pilot percentage on error probability and spectral efficiency are verified. After that, the tradeoff relation between error probability and spectral efficiency is elicited.

\begin{figure}[!t]\centering
\includegraphics[angle=0,scale=0.41]{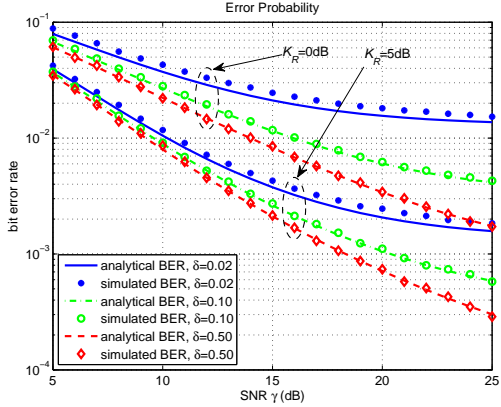}
\caption{error probability as a function of SNR $\gamma$ and pilot percentage $\delta$.}
\label{Fig_BER_SNR}
\end{figure}

Fig. \ref{Fig_BER_SNR} demonstrates the relation between $P_{e}$ and $\gamma$ under various simulation parameters, i.e., $\delta=0.02$, $0.10$, $0.50$ and $K_{R}=0$dB, $5$dB. The scattering objects spatial parameter $c_{0}=0.1\textrm{m}^{-1}$. We can see that there exists a negative relationship between $P_{e}$ and $\gamma$, $K_{R}$ as well as $\delta$, which verifies Theorem 1, Theorem 2 and the error probability expressions (\ref{BER})(\ref{Avg-BER}). Notably, it can be observed that there is an error floor for this communication system, the reason is that the MSE at data locations is independent of $\gamma$ in high SNR regime.

\begin{figure}[!t]\centering
\includegraphics[angle=0,scale=0.41]{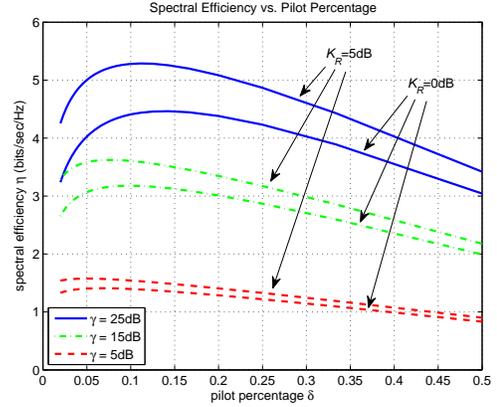}
\caption{The relation between spectral efficiency and pilot percentage.}
\label{Fig_SE_ana}
\end{figure}

Next, the relation between $\eta_{up}$ and $\delta$ is elucidated in Fig. \ref{Fig_SE_ana}, in which the spectral efficiency rises first, and then falls. In this sense, the spectral efficiency can be maximized. However, with the pilot percentage as $\delta_{\textrm{SE-opt}}$, the BER performance is less than satisfactory, especially in high SNR regime. It is practical to select a slightly higher $\delta$ which would balance between error probability and spectral efficiency.

\begin{figure}[!t]\centering
\includegraphics[angle=0,scale=0.41]{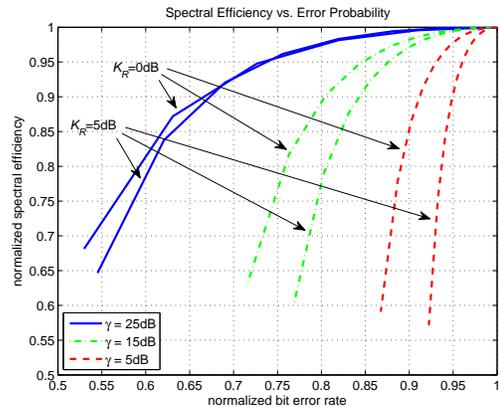}
\caption{The relation between spectral efficiency and error probability.}
\label{Fig_Tradeoff_ana}
\end{figure}

The tradeoff between error probability and spectral efficiency is depicted in Fig. \ref{Fig_Tradeoff_ana}. The feasible pilot percentages are several discrete values in range $\left[ \delta_{\textrm{SE-opt}}, 0.5 \right]$, along which $P_{e}$ and $\eta_{up}$ are both maximized when $\delta=\delta_{\textrm{SE-opt}}$. For normalization, $P_{e}$ and $\eta_{up}$ are divided by $P_{e}(\delta_{\textrm{SE-opt}})$ and $\eta_{up}(\delta_{\textrm{SE-opt}})$, respectively. From this plot, it can be seen that the tradeoff between $P_{e}$ and $\eta_{up}$ is elucidated as a polyline, the corresponding pilot percentages are $\frac{1}{2}$, $\frac{1}{3}$, \dots, $\delta_{\textrm{SE-opt}}$. Moreover, when $\gamma$ and $K_{R}$ get high, a small loss of spectral efficiency will bring in conspicuous BER performance gain.

\section{Conclusion and Future Work}

In this paper, \emph{delay correlation} inherent in HSTC system is exploited to provide robust communication under high mobility while reducing the use of pilot symbols. A novel staticized channel model based on \emph{delay correlation} is proposed and analyzed, which converts the rapid fading channel into a virtual static channel. In particular, we derived the closed-form expressions for two metrics of interest, i.e., error probability and spectral efficiency, and obtained an analytical tradeoff between them. This tradeoff provides useful references for the parameter design in future delay-correlation-enabled systems.

\emph{Delay correlation} provide an alternative robust space-time communication solution to many existing systems. In our current scheme, it is assumed that only one antenna of the receive array is selected to be active each time. We plan to extend the number of activated antennas and obtain diversity gain.

\appendix
\subsection{Proof of Theorem 1}

(\ref{MSE-Pilot-Locations}) is equal to summarizing the eigenvalues of $\frac{1}{N_{p}}\mathbf{R}_{ee}$ as follows
\begin{equation}
\small
\begin{split}
\sigma_{p,N_{p}}^{2}&=\frac{1}{N_{p}}\sum_{n=1}^{N_{p}}\left [ \lambda _{n}-\left ( \lambda _{n}+\frac{1}{\gamma } \right )^{-1} \lambda _{n}^{2} \right ]\\
&=\frac{1}{N_{p}}\sum_{n=1}^{N_{p}}\left( \frac{\lambda_{n}}{\lambda_{n}\gamma+1} \right)
\label{MSE-Pilot-Locations-Eigen}
\end{split}
\end{equation}
where $\lambda_{n}$ is the $n$-th eigenvalue of $\mathbf{R}_{hh}$. Based on the asymptotic analysis method in \cite{gazzah2001asymptotic}, when $N_{p}\rightarrow\infty$, $\sigma_{p,N_{p}}^{2}$ can be recast as
\begin{equation}
\small
\sigma_{p}^{2}=\lim_{N_{p}\rightarrow \infty }\sigma_{p,N_{p}}^{2}=\frac{1}{2\pi}\int_{-\pi}^{\pi}\left [ \frac{\Lambda\left ( \Omega  \right ) }{\Lambda\left ( \Omega  \right ) \gamma +1} \right ]d\Omega
\label{Asymptotic-MSE-Pilot-Locations-Eigen}
\end{equation}
where $\Lambda\left ( \Omega  \right )$ is the discrete-time Fourier transform (DTFT) of $\left\{\frac{1}{K_{R}+1}e^{-c_{0}\left | k \right |\frac{D}{\delta }}\right\}_{k}$, which is expressed as
\begin{equation}
\small
\begin{split}
\Lambda\left ( \Omega  \right )&=\sum_{k=-\infty }^{\infty }\left ( \frac{1}{K_{R}+1}e^{-c_{0}\left | k \right |\frac{D}{\delta }} \right )e^{-jk\Omega }\\
&=\frac{1}{K_{R}+1}\left[ \frac{1-\alpha^{2}}{ 1- 2\alpha\textrm{cos}(\Omega)+\alpha^{2} }\right]
\label{DTFT}
\end{split}
\end{equation}
where $\alpha=e^{-c_{0}\frac{D}{\delta}}$. Therefore, $\sigma_{p}^{2}$ could be calculated by
\begin{equation}
\small
\begin{split}
\sigma_{p}^{2}&=\frac{1}{2\pi}\int_{-\pi}^{\pi}\left [
\frac{\frac{1}{ K_{R} +1}}
{\frac{\gamma}{ K_{R} +1}+\frac{1+\alpha^{2}}{1-\alpha^{2}} -\frac{2\alpha}{1-\alpha^{2}}\textrm{cos}\left( \Omega \right)} \right ]d\Omega\\
&=\frac{1}{K_{R}+1}\times\sqrt{\frac{1}{a^{2}-b^{2}}}
\label{Asymptotic-MSE-Pilot-Locations-Integral}
\end{split}
\end{equation}
where $a=\frac{\gamma}{ K_{R} +1}+\frac{1+\alpha^{2}}{1-\alpha^{2}}$ and $b=-\frac{2\alpha}{1-\alpha^{2}}$.

\subsection{Proof of Theorem 2}

The same with (\ref{MSE-Pilot-Locations-Eigen}) and (\ref{Asymptotic-MSE-Pilot-Locations-Eigen}), $\sigma_{d,u,N_{p}}^{2}$ can be rewritten as
\begin{equation}
\small
\sigma_{d,u}^{2}=\frac{1}{2\pi}\int_{-\pi}^{\pi}\left [ \Lambda\left ( \Omega  \right ) - \frac{\left|\Lambda_{dh,u}\left ( \Omega  \right )\right|^{2} }{\Lambda\left ( \Omega  \right ) +\frac{1}{\gamma}} \right ]d\Omega
\label{Asymptotic-MSE-Data-Locations-Eigen}
\end{equation}
$\Lambda_{dh,u}\left ( \Omega  \right )$ is the DTFT of $\left\{\frac{1}{K_{R}+1}e^{-c_{0}\left | k +\delta u \right |\frac{D}{\delta }}\right\}_{k}$ and is expressed as
\begin{equation}
\small
\Lambda_{dh,u}\left ( \Omega  \right )=\frac{1}{K_{R}+1}\left[ \frac{\alpha\left( \beta^{-1}-\beta \right) e^{j\Omega} + \beta-\alpha^{2}\beta^{-1}}{ 1- 2\alpha\textrm{cos}(\Omega)+\alpha^{2} }\right]
\label{DTFT2}
\end{equation}
where $\beta=e^{-c_{0}uD}$. Substituting (\ref{DTFT}) and (\ref{DTFT2}) into (\ref{Asymptotic-MSE-Data-Locations-Eigen}), $\sigma_{d,u}^{2}$ could be recast as the form in (\ref{Theorem2}).

\subsection{Proof of Lemma 1}
The capacity $C( \hat{h},\gamma )$ is equivalent to the maximum conditional mutual information $\max\{\textrm{I}( x;y | \hat{h} )\}$, where $\textrm{I}( x;y | \hat{h} )$ is defined below
\begin{equation}
\small
\textrm{I}\left( x;y | \hat{h} \right)=\textrm{E}_{x,y}\left[ \log p\left( y| x,\hat{h} \right) \right]-\textrm{E}_{y}\left[ \log p\left( y| \hat{h} \right) \right]
\label{Mutual-Information}
\end{equation}
where $x$ is a symbol of the $u$-th group, $\hat{h}$ and $y$ are the corresponding estimated channel coefficient and received signal. Conditional on $\hat{h}$ and $x$, $y$ is Gaussian distributed with mean and variance given in \cite{ningsun2014spectral}
\begin{equation}
\small
\mu_{y|x,\hat{h}}=\sqrt{E_{0}}\hat{h}x
\label{y-Mean}
\end{equation}
\begin{equation}
\small
\sigma_{y|x,\hat{h}}^{2}=E_{0}\sigma_{d,u}^{2}+\sigma_{n}^{2}
\label{y-Variance}
\end{equation}
$\left| x_{u} \right|=1$ is utilized in the second equation. Therefore,
\begin{equation}
\small
\textrm{E}_{x,y}\left[ \log p\left( y| x,\hat{h} \right) \right]=\log\frac{1}{\pi e\left ( E_{0}\sigma _{d,u}^{2}+\sigma _{n}^{2} \right )}
\label{First-Item}
\end{equation}
Note that $h-\hat{h}\sim \textrm{CN}(0,\sigma_{d,u}^{2} )$, so the mean and variance of $y$ conditional on $\hat{h}$ are
\begin{equation}
\small
\mu_{y|\hat{h}}=0
\label{y-Mean2}
\end{equation}
\begin{equation}
\small
\sigma_{y|\hat{h}}^{2}=E_{0}\left (  | \hat{h}  |^{2}+\sigma _{d,u}^{2} \right )+\sigma_{n}^{2}
\label{y-Variance2}
\end{equation}
$\textrm{I}( x;y | \hat{h} )$ is maximized when $y|\hat{h}$ is Gaussian distributed. However, $y|\hat{h}$ is not Gaussian distributed in practise. In this case,
\begin{equation}
\small
-\textrm{E}_{y}\left[ \log p\left( y| \hat{h} \right) \right] \leq \log\left(\pi e\left ( E_{0}\left (  | \hat{h} |^{2}+\sigma _{d,u}^{2} \right )+\sigma_{n}^{2} \right )\right)
\label{Second-Item}
\end{equation}
Substituting (\ref{First-Item}) and (\ref{Second-Item}) into (\ref{Mutual-Information}) completes the proof.

\subsection{Proof of Theorem 3}

Since $h$ is Rician distributed with parameters $\nu=\sqrt{\frac{K_{R}}{K_{R}+1}}$ and $\sigma=\sqrt{\frac{1}{2( K_{R}+1 )}}$, i.e., $h\sim \textrm{Rice}\left(\sqrt{\frac{K_{R}}{K_{R}+1}}, \sqrt{\frac{1}{2( K_{R}+1 )}}\right)$. Hence, $|\sqrt{2(K_{R}+1)}h|^{2}$ accords with a noncentral chi-squared distribution with two degrees of freedom and noncentrality parameter $2K_{R}$.

Combining (\ref{Pilot-Location-Estimation}), (\ref{Data-MMSE-Estimation}) and (\ref{Data-Error-Correlation}), it could be obtained that
\begin{equation}
\small
\mathbf{R}_{\hat{d}\hat{d},u}=\frac{K_{R}}{K_{R}+1}\mathbf{I}_{N_{p}}+\mathbf{R}_{\hat{d}\hat{d},u}^{\textrm{DIF}}=\mathbf{I}_{N_{p}}-\mathbf{\Psi}_{ee,u}
\label{hat-h-correlation}
\end{equation}
From (\ref{hat-h-correlation}), $\textrm{E}[ |\hat{h}|^2 ]$ is given by
\begin{equation}
\small
\textrm{E}[ |\hat{h}|^2 ]=\lim_{N_{p}\rightarrow\infty}\frac{1}{N_{p}}\mathbf{R}_{\hat{d}\hat{d},u}=1-\sigma^{2}_{d,u}
\label{Mean-hat-h-squre}
\end{equation}

$|h|^{2}$ and $|\hat{h}|^{2}$ are of highly correlated distributions. Moreover, (\ref{Mean-hat-h-squre}) implies that $\textrm{E}[ |\hat{h}|^2 ]\approx\textrm{E}[ |h|^2 ]$. As a consequence, it is practical to use $|h|^{2}$ to approximate $|\hat{h}|^{2}$. Therefore,
\begin{equation}
\small
C( \hat{h},\gamma )_{u}\approx\log\left( 1+\frac{|\sqrt{2(K_{R}+1)}h|^{2}}{2\left(K_{R}+1\right)\left( \sigma_{d,u}^{2}+\frac{1}{\gamma} \right)} \right)
\label{Approximation}
\end{equation}
and
\begin{equation}
\small
\eta\approx\left( 1-\delta \right)\int_{0}^{+\infty }
\log\left( 1+\frac{x}{2\left(K_{R}+1\right)\left( \sigma_{d,u}^{2}+\frac{1}{\gamma} \right)}\right)p\left ( x \right )dx
\label{Lower-Bound}
\end{equation}
where $p\left ( x \right )$ is the probability density function of $|\sqrt{2(K_{R}+1)}h|^{2}$, which could be expressed as
\begin{equation}
\small
p\left ( x \right )=\frac{1}{2}e^{-\frac{x+2K_{R}}{2}}I_{0}\left ( \sqrt{2K_{R}x} \right )
\label{PDF}
\end{equation}
$I_{n}( z )$ is $n$-th order modified Bessel function of the first kind.

\bibliographystyle{IEEEtran}
\bibliography{bibfile}

\end{document}